Relativistic Accretion into a Reissner-Nordström Black Hole Revisited

J. A. de Freitas Pacheco

University of Nice-Sophia Antipolis, Observatoire de la Côte d'Azur
Laboratoire Cassiopée – UMR 6202
BP 4222 – 06304 Nice Cedex 4 – France
email : pacheco@oca.eu

**Abstract**

*The accretion of relativistic and non-relativistic fluids into a Reissner-Nordström black hole is revisited. The position of the critical point, the flow velocity at this point and the accretion rate are only slightly affected with respect to the Schwarzschild case when the fluid is non-relativistic. On the contrary, relativistic fluids cross the critical point always subsonically. In this case, the sonic point is located near the event horizon, which is crossed by the fluid with a velocity less than the light speed. The accretion rate of relativistic fluids by a Reissner-Nordström black hole is reduced with respect to those estimated for uncharged black holes, being about 60% less for the extreme case (charge-to-mass ratio equal to one).*

1. Introduction

The steady relativistic spherical flow of a perfect gas into a black hole has been intensely investigated in the past thirty years and a comprehensive review on the subject can be found, for instance, in [1]. For a Schwarzschild black hole, the basic relativistic equations of the inflow were discussed by Michel [2], who has derived the relations involving the sound and flow velocities at the critical point. These early investigations have shown that, for an adiabatic flow of a perfect gas with $\gamma < 5/3$ into a Schwarzschild black hole, the only critical point of the flow lying outside the horizon is that corresponding to the Bondi solution. A similar situation occurs in the case of a weakly interacting gas supposed to model dark matter. Assuming that during the inflow the phase space density is conserved, the authors in [3] derived a solution for the critical point, which is located at a distance of about 30-150 times the horizon radius for conditions expected to be present in typical dark matter halos.

Under adiabatic conditions (the cooling time is longer than the free-fall timescale), the gas is compressed as it approaches the horizon, its temperature increases and X-rays are emitted from the inner accreting envelope. This emission preheats the infalling gas, reducing considerably the accretion rate and the radiative efficiency [4]. Higher efficiencies can be obtained if dissipative turbulent motions and magnetic field line reconnection effects are included in the treatment of the inflow [5]. A consistent relativistic analysis of radiation effects on the flow were investigated in the pioneering work by Thorne and collaborators [6]. For an optically thick inflow, the distance of the critical point to the black hole horizon is reduced as well as the accretion efficiency [7]. In fact, when radiation effects are included, two branches appear in the diagram ``luminosity versus accretion rate" [8], [9] but even in the most favourable cases, corresponding to the high luminosity branch, the radiation efficiency is at maximum $2 \times 10^{-4}$, three orders of magnitude smaller than the ``canonical" value (of ten percent, derived from an accreting disk) usually adopted in cosmological simulations intended to explain the presence of supermassive black holes in the early universe by the growth of primordial seeds [10].

How the inflow properties are affected if the black hole has an electrical charge? Although the existence of charged black holes in the universe may be contested, some authors have hypothesized that such objects could play an important role in some astrophysical processes. For instance, the creation of positron-electron pairs in the ``dyadosphere'' of a Reissner-Nordström (RN) black hole was investigated in [11] (and references therein) as a possible mechanism to drive gamma-ray bursts. If the observed acceleration of the expansion of the universe is due to a phantom field, the interaction with a supermassive Schwarzschild black hole was analyzed in [12], while the interaction with a RN black hole was considered by [13] and [14]. As a consequence of this process, the mass of the black hole decreases but its charge remains constant and after a finite timescale, the black hole reaches the extreme state and may eventually produce a naked singularity [13].

As mentioned above, the formation of a charged black hole during the gravitational collapse and, in particular the formation of a ``dyadosphere'' present several difficulties whose discussion is beyond the scope of the present paper (see, however, criticisms by Page [15]). Nevertheless, the accretion process by a charged black hole offers the possibility of studying different aspects of gravity and of accretion flows in extreme conditions. In the present paper, the accretion of fluids with different equations of state into a RN black hole is revisited. We will show that for a non-relativistic baryonic fluid, the physical properties of the flow do not differ considerably from the Schwarzschild case, since the critical point is only slightly modified by the presence of a charge even in the extreme case. However, the situation is considerably different if the fluid is relativistic. In this case, the critical point is situated near the horizon and the flow velocity at this position is subsonic. Moreover, also the horizon crossing occurs at a subsonic velocity, contrary to what happens when the flow is non-relativistic, case in which the horizon is crossed with a flow velocity equal to the light velocity. No solutions for the flow were found beyond the critical point if a naked singularity is present, confirming the conclusion by [13]. For both cases, relativistic and non-relativistic flows, corrections to the accretion rate due to the presence of a charge are given. This paper is organized as follows: in Section 2 the equations of the flow are presented; in Section 3 the accretion of a non-relativistic fluid is discussed while the accretion of relativistic fluids is analyzed in Section 4. Finally, in Section 5 the main conclusions are given.

## 2. Equations of the accretion flow

The metric describing a RN black hole is given by

$$ds^2 = -B(r)dt^2 + B^{-1}(r)dr^2 + r^2 d\varpi^2 \tag{1}$$

where $d\varpi^2 = d\theta^2 + \sin^2\theta d\phi^2$ and the lapse function is defined by

$$B(r) = 1 - \frac{r_g}{r} + \frac{Q^2}{r^2} \tag{2}$$

where $r_g = 2M$ is the gravitational radius, $M$ and $Q$ are respectively the mass and the charge of the black hole. The zeros of the lapse function $B(r)$ define two horizons given by

$$r_{\pm} = \frac{r_g}{2}\left(1 \pm \sqrt{1-\beta^2}\right) \tag{3}$$

and we have defined $\beta = Q/M$. The sign "+" corresponds to the outer or to the event horizon while the sign "−" corresponds to the inner horizon. When the charge satisfies the condition $Q = M$ (or, equivalently, $\beta = 1$), both horizons coincide and this case corresponds to an extreme RN black hole.

We assume that a steady spherical inflow is set up inside the influence radius of the black hole, defined by the equality between the gravitational potential of the black hole and the mean kinetic energy of particles constituting the fluid far away from the horizon. Denoting by "$\nabla$" the covariant derivative, the conservation equations are: the mass flux conservation

$$\nabla_k J^k = 0 \tag{4}$$

where $J^k = mnu^k$ is the mass-current density. The second equation is the energy-momentum flux conservation

$$\nabla_k T_i^k = 0 \tag{5}$$

where the energy-momentum tensor is that of an ideal fluid, i.e., $T_i^k = (P+\varepsilon)u^k u_i - P\delta_i^k$, with $P$, $\varepsilon$ and $n$ being respectively the proper pressure, the proper energy density and the proper particle number density. The only non-null components of the 4-vector velocity are $u = u^1 = dr/ds$ and $u^0 = dt/ds$. In the one hand, under spherical symmetry and steady state conditions, from the time-space component of eq.5 and the expression for the stress-energy tensor one obtains

$$(P+\varepsilon)u_0 u^1 \sqrt{-g} = C_1 \tag{6}$$

where $-g$ is the metric determinant of eq.1 and $C_1$ is an arbitrary integration constant. Using the normalization condition $u_i u^i = -1$, one obtains trivially that

$$u_0 = \left(1 - \frac{r_g}{r} + \frac{Q^2}{r^2} + u^2\right)^{1/2} \tag{7}$$

On the other hand, the integration of the space component of eq 4 gives

$$nu^1 \sqrt{-g} = C_2 \tag{8}$$

where $C_2$ is another arbitrary integration constant. Substituting eq.7 into eq.6 and performing the ratio with eq.8 one obtains

$$\frac{(P+\varepsilon)}{n}\left[1 - \frac{r_g}{r} + \frac{Q^2}{r^2} + u^2\right]^{1/2} = \Delta \tag{9}$$

where $\Delta = C_1/C_2$ is a new constant. Notice that if $Q = 0$, one obtains the same result as in [2] (his eq.9) but notice that here the particle number density $n$ appears in the denominator instead of the fluid energy density. Deriving eq.9 with respect to the radial coordinate $r$ and using the mass conservation equation (eq.8), one obtains, after some algebra, the following equation

$$\frac{d \lg u}{d \lg r}\left[u^2 - V^2 F(r,u)\right] = \left[2V^2 F(r,u) - \left(\frac{r_g}{2r} - \frac{Q^2}{r^2}\right)\right] \quad (10)$$

In the above equation we have introduced respectively

$$F(r,u) = \left(1 - \frac{r_g}{r} + \frac{Q^2}{r^2} + u^2\right) \quad (11)$$

and

$$V^2 = \frac{d \lg(P+\varepsilon)}{d \lg n} - 1 \quad (12)$$

The critical point of the flow occurs when both bracketed factors in eq.10 vanish simultaneously. It should be emphasized that the critical point, contrary to usual assertions found in the literature (see, for instance [14]), does not necessarily coincide with the sonic point, in which a transition from subsonic to supersonic flow occurs. In fact, this is the situation occurring in outflows present in atmospheres of massive stars, driven by radiative forces [16, 17]. For an accreting relativistic fluid, as we shall see below, the crossing of the critical point always occurs while the flow is still subsonic.

The conditions at the critical point (coordinate $r_c$ and velocity $u_c$) are derived from the relations below, which are similar to those obtained in [14]

$$4u_c^2 = \frac{r_g}{r_c} - \frac{2Q^2}{r_c^2} \quad (13)$$

and

$$u_c^2 = V_c^2 F(r_c, u_c) \quad (14)$$

From these two equations one obtains for the critical radius

$$r_c = \frac{r_g\left(1+3V_c^2\right)}{8V_c^2}\left\{1 \pm \left[1 - \frac{8V_c^2 \beta^2\left(1+V_c^2\right)}{\left(1+3V_c^2\right)^2}\right]^{1/2}\right\} \quad (15)$$

As pointed out by [13], two solutions for the critical radius are possible. The first corresponds to the sign "+", which locates the critical radius outside the event horizon and represents a true physical solution. The other possibility corresponds to the sign "-", locates the critical radius between the inner and the outer horizon. This last solution will be discarded in the present analysis.

## 3. Accretion of a non-relativistic fluid

For a baryonic non-relativistic fluid with an equation of state $P = Kn^{\gamma}$, the energy density is given by

$$\varepsilon = mc^2 n + \frac{P}{(\gamma - 1)} \qquad (16)$$

Notice that the first term on the right side of eq.16 corresponds to the rest energy (usually neglected) while the second represents the interaction energy among the fluid particles. Using these relations and eq.12, one obtains after some algebra

$$V^2 = \frac{a^2}{\left[1 + \dfrac{a^2}{(\gamma - 1)}\right]} \qquad (17)$$

where $a^2 = \gamma P / mnc^2$ is the square of the adiabatic sound velocity measured in units of the velocity of light. Notice that solutions with $\gamma = 1$ are excluded since in this case we have a null velocity for the flow. The constant $\Delta$ in eq.9 can be calculated by imposing a zero flow velocity ($u = 0$) when $r \to \infty$ or, in other words, at distances far away from the influence sphere of the black hole. Under these conditions, one obtains trivially

$$\Delta = mc^2 \left[1 + \frac{a_{\infty}^2}{(\gamma - 1)}\right] \qquad (18)$$

where $a_{\infty}$ is the adiabatic sound velocity well beyond the influence radius of the black hole. Using this result and evaluating now eq.9 at the critical point with the help of eq.13, one obtains a relation between the adiabatic sound velocity at the critical point and that at "infinity", i.e.,

$$\frac{a_c}{a_{\infty}} \approx \sqrt{\frac{2}{(5 - 3\gamma)}} \qquad (19)$$

The flow velocity at the critical point can be evaluated from eqs.(11), (14) and (17) under the approximation $a \ll 1$ (the sound velocity is much less than the velocity of light). Performing a series expansion up to third order one obtains

$$u_c \approx a_c - \frac{3}{4}\frac{(5 - 3\gamma)}{(\gamma - 1)} a_c^3 \qquad (20)$$

Inspection of this equation reveals two distinct and important aspects: firstly, as we have mentioned before, the flow velocity at the critical point is not exactly coincident with the sound velocity, since they differ at least by a term of third order in the sound velocity.

Secondly, at the considered order, the flow velocity at the critical point is not affected by the black hole charge. However, this is not the case for the critical radius. Substituting the relations above into eq.15, one obtains for the critical radius up to second order terms in $\beta$

$$\frac{r_c}{r_g} \approx \frac{(5-3\gamma)}{8a_\infty^2}\left[1+\frac{(6-4\beta^2)a_\infty^2}{(5-3\gamma)}\right] \tag{21}$$

Notice that, as in the Schwarzschild case, the existence of a physical solution requires $\gamma < 5/3$ and that the electrical charge reduces slightly the distance of the critical point with respect to the black hole horizon.

The accretion rate can now be computed from the conditions at the critical radius, i.e.,

$$\frac{dM_{bh}}{dt} = 4\pi r_c^2 T_t^r = \pi \Gamma(\gamma)(GM_{bh})^2 f(\beta)\rho_\infty a_\infty^{-3} \tag{22}$$

In the equation above, $\rho_\infty = mn_\infty$ is the mass density of the baryonic fluid at "infinity" (beyond the influence radius of the black hole), $M_{bh}$ is the black hole mass and the functions $\Gamma(\gamma)$ and $f(\beta)$ are defined respectively as

$$\Gamma(\gamma) = \left[\frac{2}{(5-3\gamma)}\right]^{\frac{(5-3\gamma)}{2(\gamma-1)}} \tag{23}$$

and

$$f(\beta) = \left[1+\frac{(12-8\beta^2)a_\infty^2}{(5-3\gamma)}\right] \tag{24}$$

### 3.1 Some numerical results

In order to derive the radial velocity and the density profiles of the flow, let us define the dimensionless variables $x = r/r_g$ and $y = n/n_\infty$, which measure respectively the radial distance in terms of the gravitational radius $r_g$ and the particle number density in terms of its value at "infinity". Recall that the flow velocity $u$ is given in units of the velocity of light. Under these conditions, eq.9 can be written as

$$\left(1+\lambda_\infty^2 y^{\gamma-1}\right)\left(1-\frac{1}{x}+\frac{1}{4}\frac{\beta}{x^2}+u\right)^{1/2} = \left(1+\lambda_\infty^2\right) \tag{25}$$

where $\lambda_\infty^2 = a_\infty^2/(\gamma-1)$ and, as mentioned above, the adiabatic sound velocity is also given in terms of the velocity of light. Using the same notation, the particle density conservation (eq.8) can be written as

$$y = (\gamma-1)^{1/2}\lambda_\infty\left(\frac{x_c}{x}\right)^2\left(\frac{a_c}{a_\infty}\right)^{\frac{\gamma+1}{\gamma-1}} \tag{26}$$

The constant in eq.8 was calculated by applying the particle conservation equation at the critical point and using the previous result on the ratio between the adiabatic sound velocity at "infinity" and at the critical point. Equations (25) and (26) constitute an algebraic system of non-linear equations, which can be solved numerically for the fluid velocity $u$ and the ratio $n/n_\infty$ for a given value of $x = r/r_g$. The parameters characterizing the flow are the fluid temperature (or the sound velocity) at "infinity" and the adiabatic coefficient, while the parameter $\beta = Q/M$ affects the position of the critical point. The radial velocity profile of the flow (in units of the velocity of light) derived numerically from the equations above is shown in fig.1. The velocity profile was computed by assuming a fluid temperature of $10^4 K$ at "infinity" and an adiabatic coefficient $\gamma = 1.2$, adequate for an ionized gas. Solutions were obtained for different values of the charge, i.e., $\beta = 0.1$, 0.7 and 1.0 (extreme Reissner-Nordström case). For these models, the event horizon is located respectively at 0.9975, 0.8571 and 0.5000 times the gravitational radius of a Schwarzschild black hole. Notice that for all these cases, when the charge-to-mass ratio is in the range $0 < \beta \leq 1$, the flow crosses the event horizon always at the velocity of light. In these examples, the critical radius is very far from the event horizon ($r_c \approx 1.575 \times 10^8 r_g$) and the flow velocity at the critical point is about 12 $kms^{-1}$.

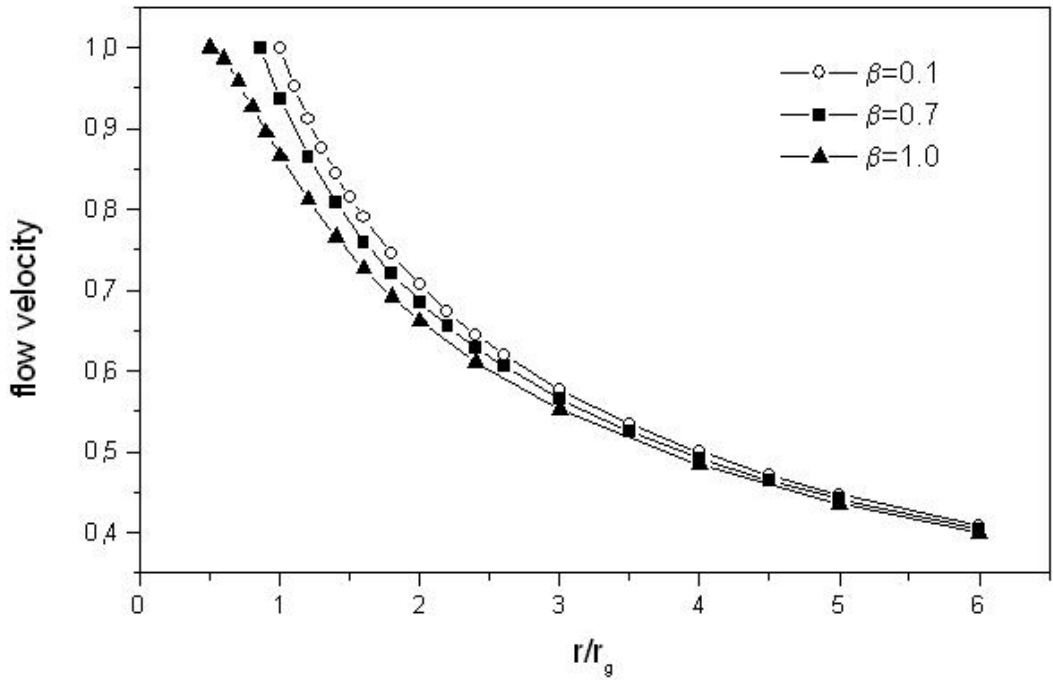

**Figure 1**. – *Radial velocity profiles for a non-relativistic fluid being accreted by a Reissner-Nordström black hole for different charge-to-mass ratios. The radial coordinate is given in terms of the gravitational radius.*

Figure 2 shows the particle density profile in terms of the density at "infinity" (compression ratio) as a function of the radial coordinate, measured in units of the gravitational radius and for different values of the charge-to-mass ratio. The compression ratio at the event horizon increases as $\beta$ increases, reaching a factor of about $2.352 \times 10^{13}$ for an extreme RN black hole. Since the temperature varies as $T \propto n^{\gamma-1}$, it may attain values of the order of $10^6 - 10^7 K$ near the horizon. It should be emphasized that these values correspond to an adiabatic flow and that radiative transfer effects may change appreciably these results.

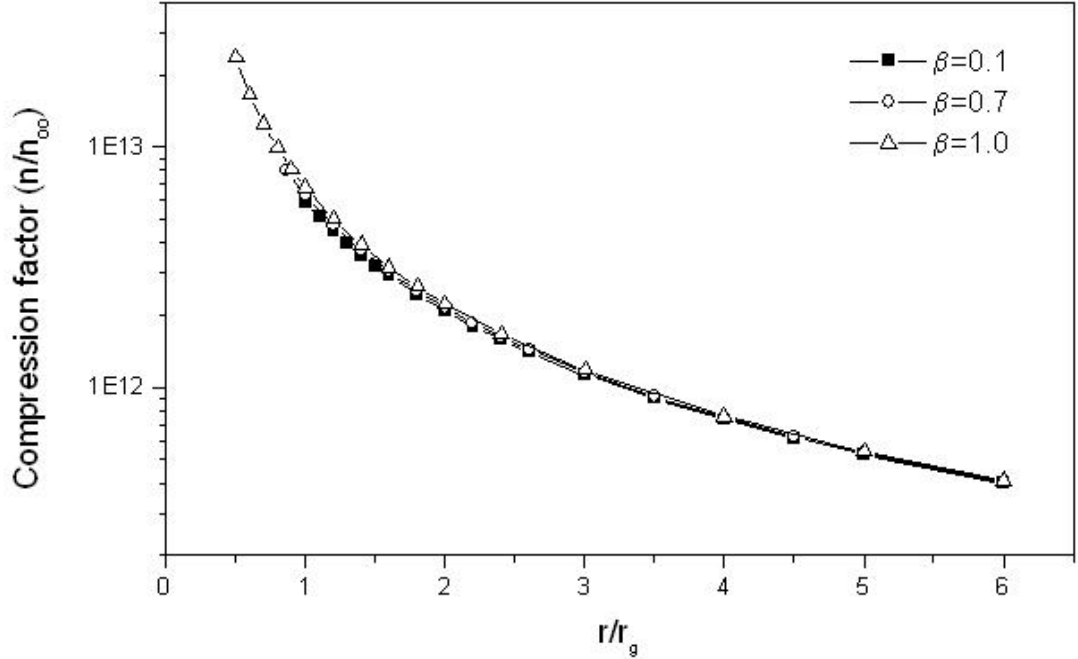

**Figure 2** – *Compression factor profiles for an inflow of a non-relativistic gas as a function of the radial coordinate in units of the gravitational radius for different charge-to-mass ratios.*

When $V_c \ll 1$ (non-relativistic flow) eq.21 admits real solutions only if the charge-to-mass ratio satisfies the condition

$$\beta^2 \leq \frac{(1+3V_c^2)^2}{8V_c^2(1+V_c^2)} \approx \frac{1}{8V_c^2} \qquad (27)$$

The extreme case $\beta^2 \approx 1/(8V_c^2) \gg 1$ corresponds to a situation of a naked singularity. The critical radius is located at $r_c \approx r_g/(8V_c^2)$, close to the sonic point within the considered approximation. However, no solution for the flow beyond the critical point was found under these circumstances. In [13] the authors suggested that when $\beta > 1$ an ideal fluid does not accrete at all onto a naked singularity, instead a static "atmosphere" will be formed. In this case, the fluid should obey the hydrostatic equilibrium equation

$$\frac{\partial P}{\partial r}+(P+\varepsilon)\frac{\partial \lg\sqrt{B(r)}}{\partial r}=0 \qquad (28)$$

The solution of this equation indicates that the matter density increases as the singularity is approached, attains a maximum and then decreases toward the limit $\rho \to 0$ as $r \to 0$. This configuration, including a density inversion is clearly unstable (Rayleigh-Taylor instability). The observed behaviour can be explained by the fact that the effective acceleration $g_{ef}=\partial \lg\sqrt{B(r)}/\partial r$ is always negative outside the horizon but becomes positive for $r<0.5\beta^2 r_g$. The repulsive force due to the modification of the space curvature near the singularity is a consequence of the presence of the electric charge. Past investigations [18] on the gravitational collapse of a uniform sphere constituted by charged dust led to a similar conclusion. Once the horizon is crossed by the surface of the sphere, the gravitational attraction is replaced by a repulsive force and, consequently, the surface of the sphere never reaches the singularity. Therefore, in a certain sense, the singularity is "protected" by the repulsive force induced by the electric charge.

## 4. Accretion of a relativistic fluid

In the case of a relativistic fluid, the equation of state is simply given by $P=\varepsilon/3$ and the pressure is related to the particle number density by $P \propto n^{4/3}$. Under these conditions, from eq.12, one obtains trivially that $V^2=1/3$. Replacing this result into eq.15 one obtains for the critical radius

$$r_c = \frac{3}{4}r_g\left[1\pm\left(1-\frac{8}{9}\beta^2\right)^{1/2}\right] \qquad (29)$$

This relation indicates that a critical point exists only if $\beta \leq 3/\sqrt{8}$. This leaves open the possibility for the existence of relativistic flows in the presence of a naked singularity. However, even in this case, no solutions were found when $\beta > 1$ for the same reasons already mentioned.

When $\beta=0$ (the case of a Schwarzschild black hole), contrary to what happens for a non-relativistic fluid, the critical radius is located near the horizon, i.e., $r_c/r_g=3/2$ and the radial velocity of the flow at this point is $u_c=1/\sqrt{6}$, implying that the crossing occurs subsonically. The flow becomes supersonic only at $r_s \approx 1.07754 r_g$, quite close the horizon, which is crossed with a velocity of $\sim 0.62c$. Using eq.9 and the conditions at the critical point, the ratio between the density or the temperature ($T \propto n^{1/3}$) at this point and the value at "infinity" can be easily computed, e.g.,

$$\left(\frac{n_c}{n_\infty}\right)^{1/3}=\left(\frac{T_c}{T_\infty}\right)=\sqrt{2} \qquad (30)$$

When $\beta \ll 1$, the critical radius is given approximately by

$$r_c \approx \frac{3}{2} r_g \left(1 - \frac{2}{9}\beta^2\right) \qquad (31)$$

Using this result and eq.13, the flow velocity at the critical point is

$$u_c^2 \approx \frac{1}{6}\left(1 - \frac{1}{9}\beta^2\right) \qquad (32)$$

Notice that the flow velocity at the critical point is reduced by the effect of the charge and the critical radius approaches the horizon.

For the extreme case ($\beta = 1$) two solutions for the critical radius exist. The first locates the critical radius at a distance twice the horizon and the flow velocity at this point is $u_c = 1/\sqrt{8}$. The other solution is not physically acceptable, since the critical radius coincides with the horizon, resulting in a null flow velocity at this position. The velocity and the density profiles of the flow can be derived as explained in Section 3.1, taking into account that for a relativistic fluid $(P+\varepsilon)/n \propto n^{1/3}$. In fig.3 the radial velocity profile is shown as a function of the radial coordinate given in units of the gravitational radius for an extreme RN black hole. After crossing the critical point, the flow reaches a maximum velocity of about $u_{max} \approx 0.4367$ at $r_{max} \approx 0.60 r_g$. Then the velocity decreases and the flow crosses the horizon with a subsonic velocity of about 0.4029c. Compression ratios attained by a relativistic fluid are always considerably smaller than those derived for the non-relativistic case. For an extreme RN black hole, the compression factor near the horizon is about 15.3, about twelve orders of magnitude less than that obtained for a non-relativistic fluid. Since in the relativistic case the temperature varies as $T \propto n^{1/3}$, near the horizon, due to the adiabatic compression, the temperature is only about 2.48 times the value at "infinity".

The accretion rate can be computed from the flow conditions at the critical point by following the same steps as before. In this case one obtains

$$\frac{dM_{bh}}{dt} = 32\sqrt{3}\pi \frac{(GM_{bh})^2}{c^5} \varepsilon_\infty \left(1 + O(\beta^4)\right) \qquad (33)$$

where $\varepsilon_\infty$ is the energy density of the fluid at "infinity". This relation differs from the Schwarzschild case only in terms of the order of $\beta^4$. It is worth mentioning that the accretion rate of relativistic particles by a black hole is usually computed by using the capture cross-section $\sigma_{cap} = 27\pi r_g^2 / 4$, leading to a numerical factor in eq.33 equal to 27 instead of $32\sqrt{3}$ here obtained. The simple use of the capture cross-section neglects hydrodynamical and relativistic effects that increase the accretion efficiency by about a factor of two. In the case of an extreme RN black hole, the accretion rate is given by

$$\frac{dM_{bh}}{dt} = \frac{512\sqrt{3}\pi}{27} \frac{(GM_{bh})^2}{c^5} \varepsilon_\infty \qquad (34)$$

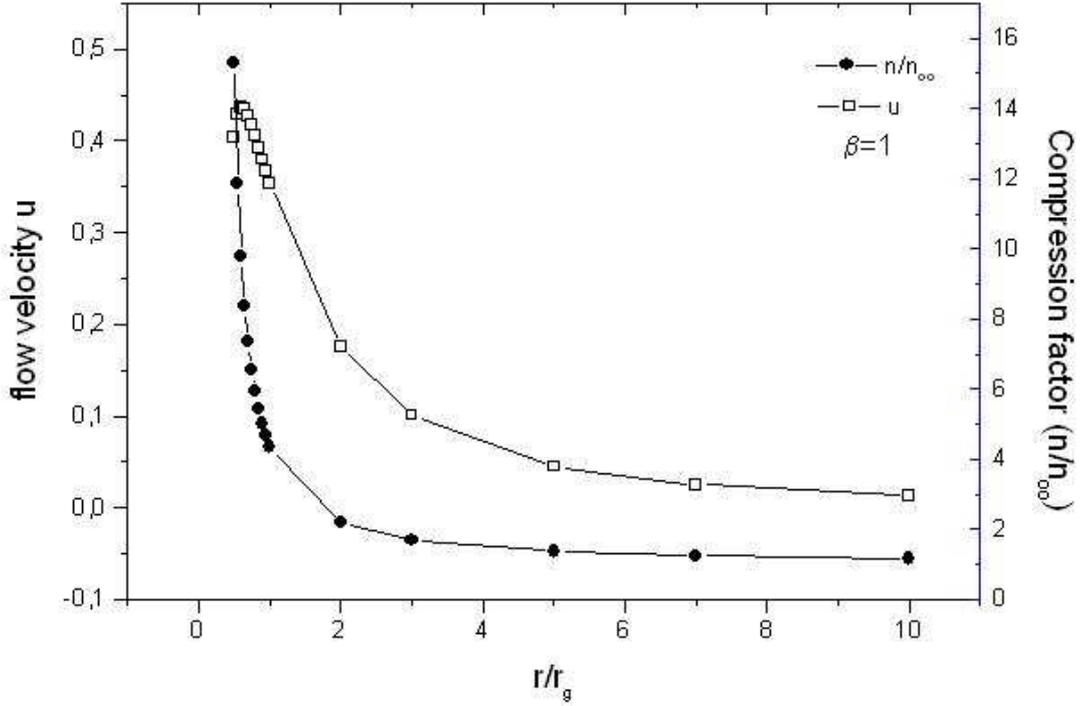

**Figure 3** – *The radial velocity profile (left ordinate) for an accreting relativistic fluid as a function of the radial coordinate, in units of the gravitational radius, for an extreme RN black hole. The profile of the compression ratio (right ordinate) is also shown.*

It can be easily verified that, in this case the accretion efficiency decreases by almost 60% with respect to an uncharged black hole.

## 5. Conclusions

The steady and spherically symmetric accretion of non-relativistic and relativistic fluids into a Reissner-Nordström black hole was revisited. A charged black hole modifies slightly the properties of the inflow of a non-relativistic fluid since the critical radius, the radial velocity at the critical point and the accretion rate are affected only by second order terms in the charge-to-mass ratio.

The situation is rather different for the inflow of relativistic fluids. Firstly, the critical point occurs closer the horizon, what is not the case for the inflow of non-relativistic fluids. Secondly, the crossing of the critical point occurs always in a subsonic regime, even if the black hole is uncharged, contrary to what is usually stated in the literature. The sonic point is reached only very near the horizon and the fluid crosses the horizon with a velocity less than the light speed. In non-relativistic flows the ratio between the particle density near the horizon and at "infinity" may attain values of the order of $10^{13}$, if the flow is adiabatic. This is not the case when relativistic flows are considered, since in this situation the compression ratio varies from about 2.83 for a weakly charged black hole up to 4.35 for the extreme case ($\beta = 1$). Finally, the accretion rate of a relativistic fluid by an extreme RN black hole is about 60% smaller than that expected for a Schwarzschild black hole.

When $\beta > 1$ the singularity is "naked" and, depending on the value of the charge-to-mass ratio, mathematical solutions for the critical point exist either for non-relativistic or for relativistic fluids. However no solutions for the flow beyond the critical point were found. Some authors [13] suggested that beyond that point a "static" solution is possible. These solutions present an inversion in the mass density profile, consequence of a repulsive force due to the black hole charge, which is manifested for distances less than $\beta^2 GM_{bh}/c^2$. Such an inversion observed in the mass density profile is probably unstable against the Rayleigh-Taylor instability, suggesting that such static solutions cannot exist.